# An Exactly Solvable Toy Model of Autocatalysis: Irreversible Relaxation after a Quantum Quench

R. Merlin

*The Harrison M. Randall Laboratory of Physics, University of Michigan,
Ann Arbor, Michigan 48109-1040, USA
and Max Planck Institute for the Structure and Dynamics of Matter,
Luruper Chausse 149, 22761 Hamburg, Germany*

A fully resolvable quantum many-body Hamiltonian is introduced that mimics the behavior of the autocatalytic chemical reaction $A + B \rightleftarrows 2B$ involving two different molecular species, *A* and *B*. The model also describes two nonlinearly-coupled modes of an optical cavity. Consistent with the current understanding of the relaxation dynamics of integrable systems in isolation, the wavefunction following a quantum quench exhibits irreversibility with retention of the memory about its initial conditions. Salient features of the model include a marked similarity with conventional quantum decay and a total *B*-to-*A* conversion, with associated classical-like behavior of the wavefunction, when the initial state does not contain *A*-type molecules.

Autocatalysis refers to chemical reactions in which a compound serves as the catalyzer of its own formation. Such reactions are found in a wide range of phenomena and problems of interest to many disciplines, from chemistry and biology to materials synthesis and environmental science [1, 2]. Notable examples include asexual reproduction [3], drug transport [4], prion replication [5], prebiotic chemistry [6] and molecular self-assembly [7], among many others. For the most part, theoretical models of autocatalysis are of classical nature. They usually involve a set of non-linear deterministic or stochastic rate equations [8] whose solutions, leading to, e. g., chemical oscillations, spontaneous pattern formation or chaos [9], cover a broad collection of complex dynamical behavior. Here, an exactly solvable quantum toy model of autocatalysis is described, which is also an idealized model of two-mode nonlinear coupling in cavity QED [10] and Bose-Einstein condensates involving multiple species [11]. Its dynamics in the absence of a reservoir, and after a quantum quench, exhibits non-thermal irreversible behavior and other features that are not present in classical approaches. Our model adds to the relatively small group of many-body integrable problems, the physical relevance of which has increased significantly in recent years because of remarkable advances in ultracold gas experiments and the understanding of quantum thermalization [12].

The simplest autocatalytic reaction is the process [13]

$$A + B \rightleftarrows 2B \quad , \tag{1}$$

involving two molecular species, *A* and *B*, which converts a molecule of type *A* into one of type *B* (or vice versa), with a second *B* molecule acting as the catalyst. To model it, we assume the molecules to be bosonic particles, which interact with each other according to the Hamiltonian [14]

$$\hat{H} = -\frac{\Lambda}{N}(b^{\dagger}a^{\dagger}bb + b^{\dagger}b^{\dagger}ab) \quad . \tag{2}$$

Here, $a$ and $a^\dagger$ ($b$ and $b^\dagger$) are respectively the annihilation and the creation operator for the $A$ ($B$) bosons, $N$ is the total number of particles and $\Lambda > 0$. We note that related Hamiltonians have been considered to model two-species Bose–Einstein condensates and two modes of the electromagnetic field interacting in a Kerr-media cavity [15,16,17].

Let $|n_A, n_B\rangle$ be the state containing exactly $n_A$ and $n_B$ molecules of type $A$ and $B$. Given that the Hamiltonian commutes with the total number of particles and that $|n_A = N, n_B = 0\rangle$ is an eigenstate, of zero energy, the remaining eigenstates are of the form $|\Psi_E\rangle = e^{-iEt}|\Phi_E\rangle$ with

$$|\Phi_E\rangle = \sum_{j=1}^{N} \Xi_j^E |N-j, j\rangle \quad , \tag{3}$$

where $\Xi_j^E$ are expansion coefficients, $E$ is the energy and $t$ is the time ($\hbar = 1$). From Schrödinger equation, we have

$$\Xi_{j+1}^E \sqrt{(N-j)(j+1)j^2} + \Xi_{j-1}^E \sqrt{(N-j+1)j(j-1)^2} = -\frac{EN}{\Lambda} \Xi_j^E \quad . \tag{4}$$

In the limit $N-j, j \gg 1$, the above simplifies to

$$\Xi_{j+1}^E + \Xi_{j-1}^E \approx -\frac{EN}{\Lambda \sqrt{(N-j)j^3}} \Xi_j^E \quad . \tag{5}$$

It follows that the limiting spectrum is symmetric with respect to $E = 0$ in that if $\{\Xi_j^{\tilde{E}}\}$ is the set associated with a particular eigenvalue $\tilde{E}$, then the set $\{(-1)^{j+1}\Xi_j^{\tilde{E}}\}$ is also a solution, with eigenvalue $-\tilde{E}$. For $N \gg 1$, and provided $E \leq 0$, calculations show that one can treat the discrete set of coefficients in Eq. (3) as a continuous-variable function $\Xi^E(c)$ so that Eq. (5) becomes [18]

$$\left[ -\frac{1}{2N} \frac{d^2}{dc^2} + U_E(c) \right] \Xi^E = 0 \quad , \tag{6}$$

where $c = j/N$ is the concentration of $B$ molecules and

$$U_E(c) = -N\left(1 + \frac{E}{2N\Lambda\sqrt{(1-c)c^3}}\right) \quad . \tag{7}$$

This expression is identical to the Schrödinger equation of a particle of mass $N$ and zero energy moving in a gravitational-like potential [14]. For $E \leq 0$, $U_E(c)$ has a minimum at $c_m = ¾$. Setting $U_E(c_m) = 0$, we obtain the ground-state energy $E_{GS} = -3\sqrt{3}\Lambda N/8$. Hence, the energy spectrum is confined to the range

$$-\frac{3\sqrt{3}}{8}\Lambda N < E < \frac{3\sqrt{3}}{8}\Lambda N \quad . \tag{8}$$

Since the potential in the vicinity of $c_m$ is that of a harmonic oscillator, we expect the spectrum near the boundaries to exhibit a ladder of equally spaced levels. A simple calculation gives the (adimensional) oscillator frequency $\Omega = \sqrt{32/3}$. To find the level spacing, we write $E = E_{GS} + \delta$ in Eq. (6) and, after expanding $U_E$, we obtain $\delta = 3\Lambda(n+1/2)/\sqrt{2}$, with integer $n$. The procedure also gives the ground state wavefunction

$$|\Psi_{GS}\rangle = \left(\frac{\Omega N}{\pi}\right)^{1/4} e^{-iE_{GS}t} e^{-\Omega N(c-c_m)^2/2} \quad , \tag{9}$$

which is highly localized in concentration space, as are all the oscillator-related eigenstates at the top and bottom of the spectrum.

Unlike states near the spectral boundaries, numerical results show that a generic wavefunction extends throughout a wide range of concentrations and, moreover, that the eigenenergies in the vicinity of $E = 0$ $\qquad \approx \pm \Lambda\sqrt{8p^3/3N}$ $\Xi^E(c) = e^{if_E(c)}$, we get $df_E/dc \approx \sqrt{-2NU_E}$. For a given $E < 0$, the allowed concentration range is defined by the classical turning points, $c_1$ and $c_2$, at which $U_E = 0$, whereas approximate values of the eigenenergies can be obtained from the Bohr–Sommerfeld condition $f_E(c_2) - f_E(c_1) = k\pi$, with $k = N/2 - p$ [14]. Consistent with Eq. (8), $U_E$ becomes positive everywhere and, thus, there are no solutions for $E < -3\sqrt{3}\Lambda N/8$.

If our system of particles is in contact with an appropriate bath, general principles of statistical mechanics indicate that, regardless of the initial conditions, it will evolve into a state of thermal equilibrium involving the lowest-lying states localized at $c = c_m$. This means that the equilibrium constant for the chemical reaction of Eq. (1) is

$$K = \frac{[A]}{[B]} = \frac{1 - c_m}{c_m} = 1/3 \quad , \tag{10}$$

where the brackets denote the average concentration. In the following, we assume our system to be completely isolated from the environment and focus on its coherent dynamics, with time evolution dictated by the Schrödinger equation. Since an arbitrary initial state can be written as

$$\sum_{j=0}^{N} C_j |N-j, j\rangle \quad , \tag{11}$$

without loss of generality, we can confine our analysis to studying the dynamics of initial states with a fixed number of $A$ and $B$ molecules, that is, $|N - j_0, j_0\rangle$, for which the expectation value of the energy is exactly zero. Seeing that such states are stationary solutions for $\Lambda = 0$, this process represents a quantum quench [19,20] involving the abrupt turn-on of the coupling constant $\Lambda$.

For $t > 0$, the wavefunction of our system after the quench is

$$|\Psi\rangle = \sum_E |\Phi_E\rangle \langle \Phi_E | N - j_0, j_0 \rangle e^{-iEt} = \sum_E \tilde{\Xi}^E_{j_0} |\Phi_E\rangle e^{-iEt} \tag{12}$$

(the tilde denotes the complex conjugate). Therefore, the time-dependent concentration of B molecules is $\langle c(j_0,t)\rangle = \sum_j jP(j,j_0,t)/N$ where

$$P(j,j_0,t) = \left|\sum_E \tilde{\Xi}^E_{j_0} \Xi^E_j e^{-iEt}\right|^2 \qquad (13)$$

is the probability of finding the system in the state $|N-j,j\rangle$ at time $t$. For large $N$, numerical studies reveal that $\langle c(j_0,t)\rangle$ reaches a steady-state that depends on the initial state concentration $c_0 = j_0/N$ in times on the order of $\Lambda^{-1}$. In Fig. 1, we plot the final average concentration, that is, $\lim_{t\to\infty}\langle c(j_0,t)\rangle$, as a function of $c_0$. Interestingly, it peaks at $c = c_m$, as does the ground state wavefunction. The steady-state values do not depend much on the total number of particles except for $c_0 \approx 1$ where it behaves $\propto N^{-\alpha}$, as shown in the inset for $c_0 = 1$. Hence, we expect there a complete conversion from B- into A-type molecules for $N \to \infty$.

Results of computational simulations for $c_0 = 1$ are illustrated in Fig. 2. The time-behavior of the wavefunction is reminiscent of that of a quasi-classical particle, with the concentration playing the role of position. With increasing time, the wavefunction first broadens before narrowing down until it reaches a minimum value beyond which the state spreads throughout the whole range of concentration. The time-dependence shown in the inset exhibits a striking resemble to quantum decay in that, after a short time and up to the minimum at $t \approx 3.6$, the average concentration decays exponentially.

The numerical result that the average concentration approaches constant values for large $N$ and $t \gg 1/\Lambda$ is a strong indication of the quantum irreversibility of our model. Closed macroscopic quantum systems are generally believed to either keep information about their initial conditions, as is the case of systems that exhibit many-body localization [21,22], or to thermalize according to a conventional [19] or, for integrable problems, generalized Gibbs distribution [23,24,25]. Our autocatalytic model, exhibiting both relaxation and memory of the initial conditions, belongs to the latter group.

The source of irreversible behavior in our system is the cancellation of off-diagonal terms in the expression for the probability so that

$$P(j,j_0,\infty) = \lim_{t\to\infty} \sum_{E,E'} \tilde{\Xi}^E_{j_0} \Xi^{E'}_{j_0} \Xi^E_j \tilde{\Xi}^{E'}_j e^{-i(E-E')t} = \sum_E \left|\Xi^E_{j_0} \Xi^E_j\right|^2 \quad (14)$$

and, therefore,

$$\langle c(j_0,\infty)\rangle = \sum_{E,j} j \left|\Xi^E_{j_0} \Xi^E_j\right|^2 / N \quad . \quad (15)$$

Using that $\Xi^E_j = (-1)^{j+1} \Xi^{-E}_j$, the WKB approximation and the method of steepest descent, the cancellation follows from the fact that the integral

$$\int_{-E_{GS}}^{0} e^{i[f_E(c_0)-f_E(c)-Et]} dE \quad (16)$$

vanishes for $t \to \infty$ since $df_E/dE < \infty$. Provided $C_j$ in Eq. (11) is a slowly-varying function of $j$, it is apparent that the above arguments apply as well to arbitrary initial states. More generally, the long-term behavior of a generic state

$$\sum_E B_E |\Phi_E\rangle e^{-iEt} \quad (17)$$

vis-à-vis the concentration is determined by the correlation

$$F(E,E') = \sum_j j \Xi^E_j \tilde{\Xi}^{E'}_j / N \quad . \quad (18)$$

Given that $\sum_j \Xi^E_j \tilde{\Xi}^{E'}_j = \delta_{E,E'}$  $N, t \to \infty$  $E \neq E'$ and, thus, $\langle c(t)\rangle$ should reach a constant value unless the two energies are closely spaced. Since the energy separation between neighboring states near $E = 0$ as well as the width of the states at the bottom and top of the spectrum scale like $1/\sqrt{N}$, oscillations with a period of order $1/\Lambda$ can only occur at intermediate concentrations.

As shown in Fig. 1, the expression for the steady-state concentration, Eq. (15), agrees extremely well with the time-domain calculations. Moreover, the long-term probability plots of Fig. 3, showing prominent peaks at $c_0$, clearly illustrate the dependence of the relaxed state on the initial conditions.

In conclusion, we described an exactly solvable quantum toy model of the chemical reaction $A+B \rightleftarrows 2B$, whose dynamics differs significantly from that of classical rate equations. In the particular case where the initial state has no molecules of type $A$, the system evolves semi-classically into a state without molecules of type $B$, following a time behavior similar to that of quantum decay. After a quantum quench, the wavefunction exhibits irreversible behavior with steady states that depend on the initial conditions, as expected for integrable Hamiltonians. These results add to the relatively small set of exact solutions to many-body problems that can help elucidate important questions about the dynamics of two-coupled modes in cavity QED and two-species Bose-Einstein condensates, as well as on the quantum thermalization of isolated systems.

Stimulating discussions with D. Jaksch are gratefully acknowledged.

# REFERENCES

[1] P. Schuster, Monatsh. Chem. **150**, 763 (2009).

[2] A. J. Bissette and S. P. Fletcher, Angew. Chem. Int. Ed. **52**, 12800 (2013).

[3] K. E. Holsinger, Proc. Natl. Acad. Sci. U.S.A. **97**, 7037 (2000).

[4] A. N. Ford Versypt, D. W. Pack and R. D. Braatz, J. Control Release **165**, 29 (2013).

[5] J. Bieschke, P. Weber, N. Sarafoff, M. Beekes, A. Giese and H. Kretzschmar, Proc. Natl. Acad. Sci. U.S.A. **101**, 12207 (2004).

[6] M. Preiner, J. C. Xavier, A. do Nascimento Vieira, K. Kleinermanns, J. F. Allen and W. F. Martin, Interface Focus **9**, 20190072 (2019).

[7] G. Ashkenasy, T. M. Hermans, S. Otto and A. F. Taylor, Chem. Soc. Rev. **46**, 2543 (2017).

[8] R. Plasson, A. Brandenburg, L. Jullien and H. Bersini, J. Phys. Chem. A **115**, 8073 (2011).

[9] I. R. Epstein and J. A. Pojman (Oxford University Press, New York, 1998).

[10] S. Franke, M. Richter, J. Ren, A. Knorr and S. Hughes, Phys. Rev. Res. **2**, 033456 (2020).

[11] L.-M. Duan, J. I. Cirac and P. Zoller, Nature **409**, 63 (2001).

[12] M. Rigol, V. Dunjko, V. Yurovsky and M. Olshanii, Phys. Rev. Lett. **98**, 050405 (2007).

[13] E. Arslana and I. J. Laurenzib, J. Chem. Phys. 128, 015101 (2008).

[14] R. Merlin, Europhys. Lett., **76**, 541 (2006).

[15] A. B. Klimov and L. L. Sanchez-Soto, Phys. Rev. A **61**, 063802 (2000).

[16] L. Sanz, R. M. Angelo and K. Furuya, J. Phys. A: Math. Gen. **36**, 9737 (2003).

[17] S. B. Ocak, Ö. Yeşiltaş and B. Demircioğlu, Int. J. Theor. Phys. **47**, 1865 (2008).

[18] For $\Lambda < 0$, $\Xi_j^E$ can be replaced with a continuous function for positive energies.

[19] L. D'Alessio, Y. Kafri, A. Polkovnikov, and M. Rigol, Adv. Phys. **65**, 239 (2016).

[20] A. Mitra, Annu. Rev. Condens. Matter Phys. **9**, 245 (2018).

[21] R. Nandkishore and D. A. Huse, Annu. Rev. Condens. Matter Phys. **6**, 15 (2015).

[22] D. A. Abanin, E. Altman, I. Bloch, and M. Serbyn, Rev. Mod. Phys. **91**, 021001 (2019).

[23] A. C. Cassidy, C. W. Clark, and M. Rigol, Phys. Rev. Lett. **106**, 140405 (2011).

[24] J. S. Caux and F. H. L. Essler, Phys. Rev. Lett. 110, 257203 (2013).

[25] E. Ilievski, J. De Nardis, B. Wouters, J.-S. Caux, F. H. L. Essler and T. Prosen, Phys. Rev. Lett. **115**, 157201 (2015).

# FIGURE CAPTIONS

**Figure 1** - Dependence of the steady-state average concentration of $B$ particles on the initial-state concentration $c_0 = j_0/N$, for $\Lambda = 1$ and various values of $N$. The curve is the theoretical expression, Eq. (15), for $N = 180$ and the arrow labels $c = c_m$. The inset shows a log-log plot of the dependence on $N$ for $c_0 = 1$. The line though the points is $\propto N^{-\alpha}$ with $\alpha \approx 0.157$.

**Figure 2** – Time evolution of the probability of finding the system at a given concentration of $B$ particles. From right to left, $t = 0.2, 0.85, 1.5, 2.15, 2.8$ and $3.45$ (note the equal time intervals). The initial state is $|0, N\rangle$. The time-dependence of the average concentration, $\langle c(t) \rangle$, is shown in the inset. Data for $\Lambda = 1$ and $N = 180$. After reaching the minimum value, the wavefunction develops a high-concentration tail and eventually spreads throughout the whole range of concentrations.

**Figure 3** – Dependence of the steady-state probability of finding the system in the state $|N-j, j\rangle$ after a quench from $|N-j_0, j_0\rangle$ for $N = 200$, $\Lambda = 1$ and different values of $c_0 = j_0/N$; see Eq. (14). The abscissa is the concentration $c = j/N$. Arrows denote the narrow features at $c = c_0$.

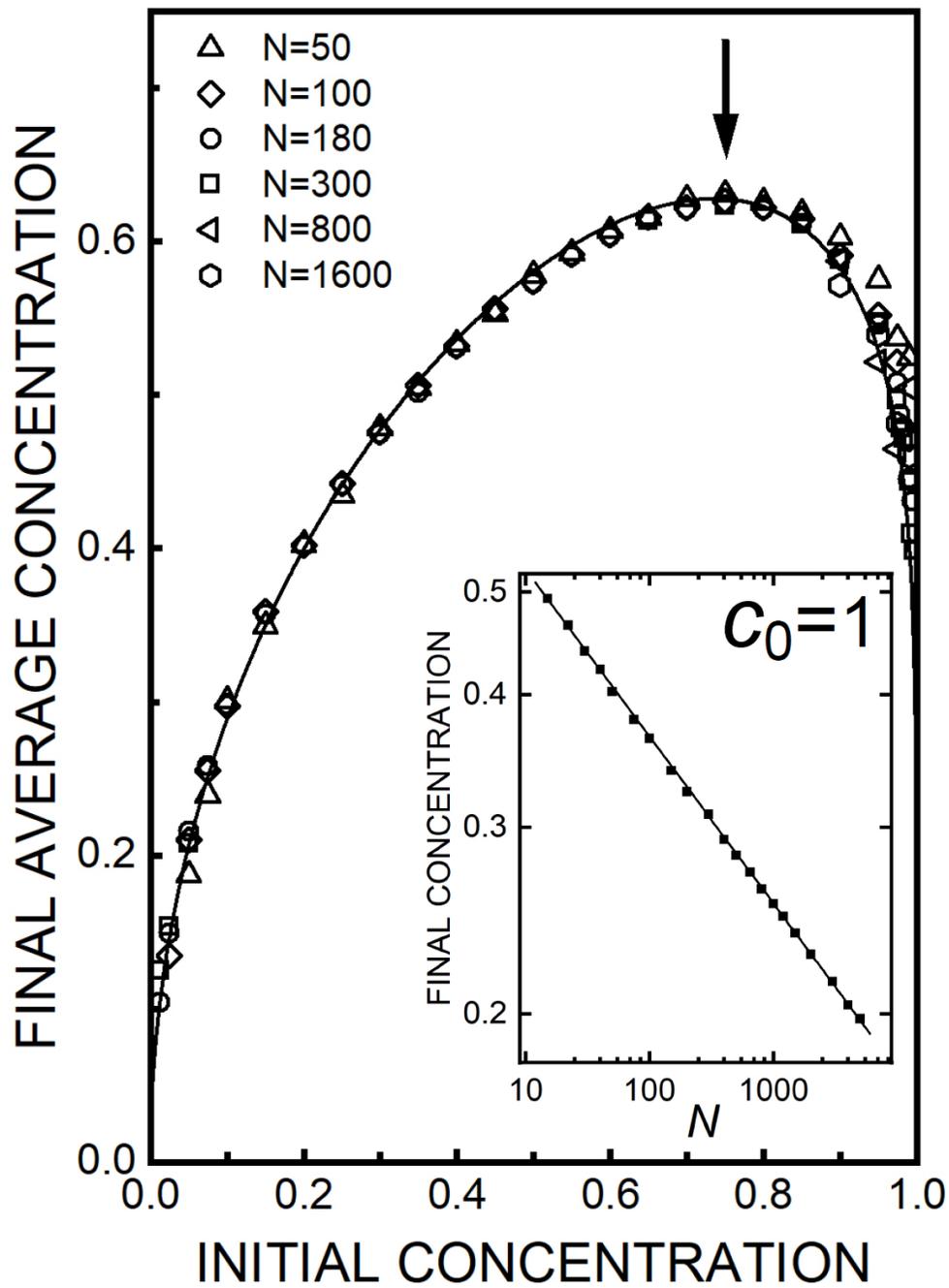

FIGURE 1

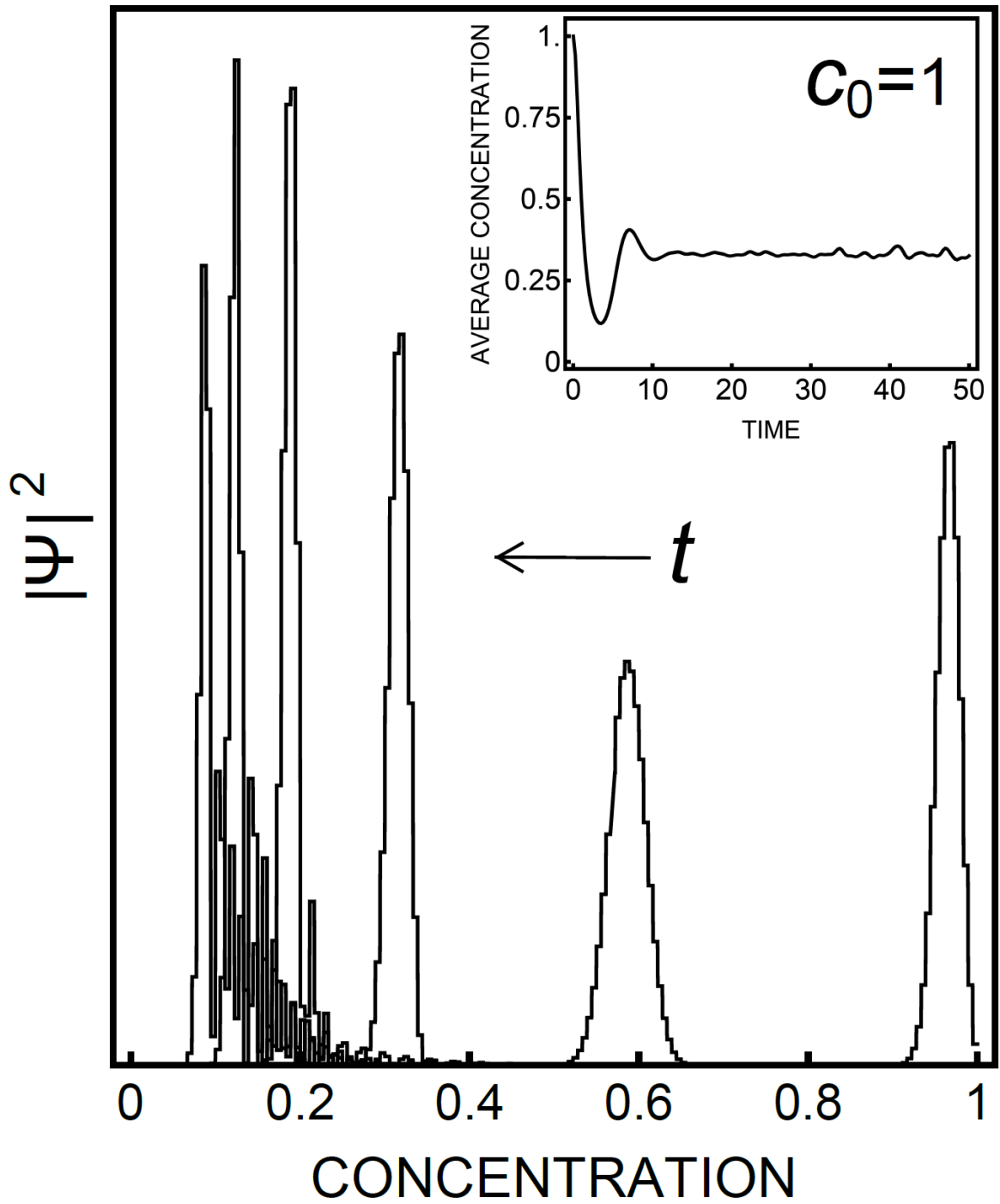

FIGURE 2

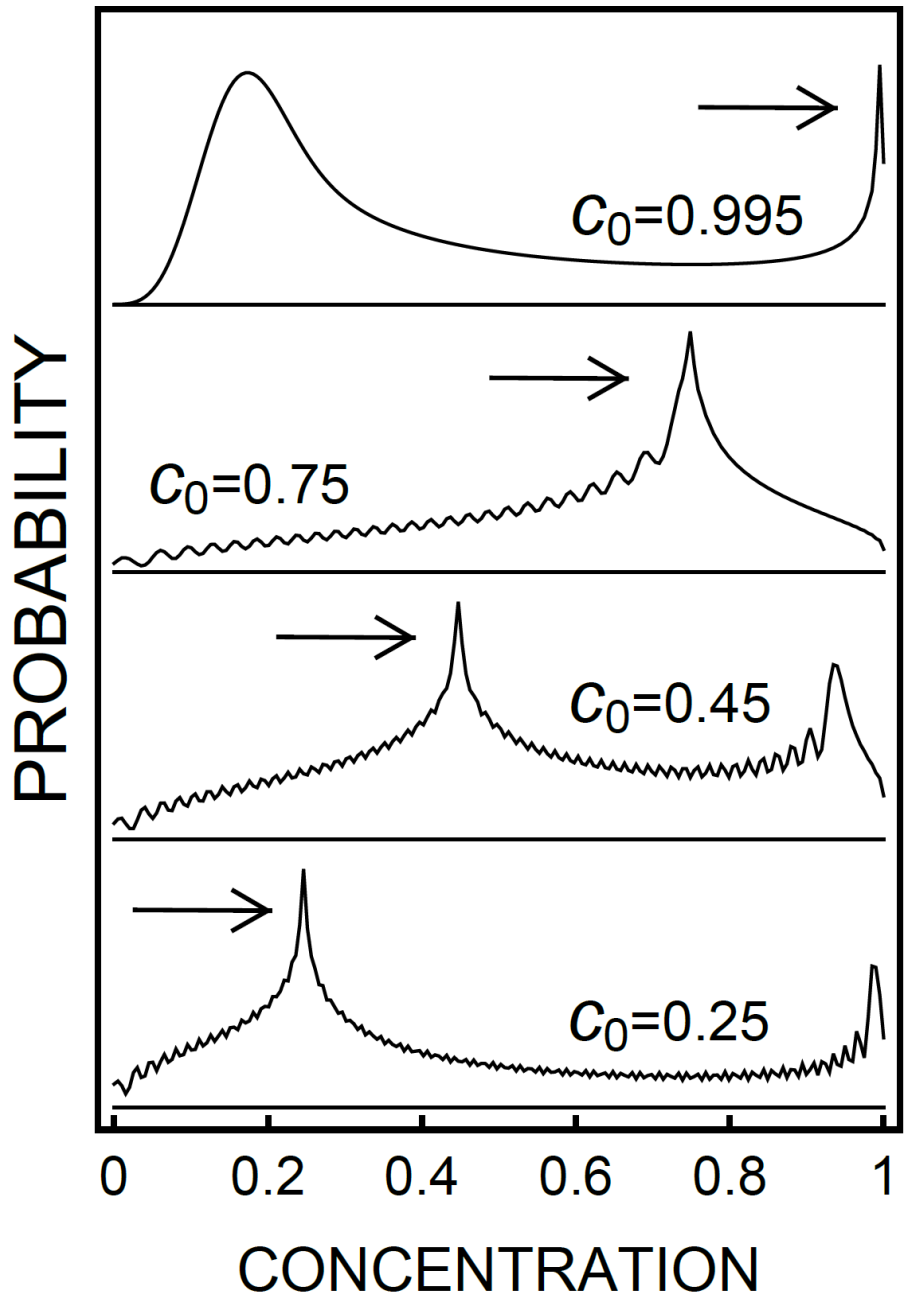

FIGURE 3